\documentclass[prl,twocolumn]{revtex4}
\usepackage{bm}

\newfam\mbffam 

\font\tenscr=rsfs10

\font\twelvescr=rsfs10 scaled \magstep 1
\skewchar\tenscr='177
\newfam\scrfam \textfont\scrfam=\twelvescr

\def\avg#1{\langle#1\rangle}
\def\Re{\mbox{Re}}
\def\Im{\mbox{Im}}
\def\Vk{{\bm k}} \def\Vq{{\bm q}} \def\VQ{{\bm Q}}
\def\Vx{{\bm x}} \def\Vy{{\bm y}} \def\V0{{\bm 0}}

\begin{document}
\title{\Large\bf  Doping induced spinless collective excitations of charge
$2e$ in doped Mott insulators}

\author{W. Vincent Liu} 
\email[E-mail:~ ]{wvliu@uiuc.edu}
\affiliation{%
Department of Physics, University of Illinois at Urbana-Champaign,
1110 West Green Street, Urbana, Illinois 61801
}%

\date{\today}%

\begin{abstract}

Based on the Hubbard model description of the doped Mott
insulator, we prove that there
must exist a  spinless collective excitation
of charge $2e$ and of  energy $U-2\mu$ peaked at
momentum $(\pi/a,\pi/a)$ in the non-superconducting state due to doping.
Such a collective excitation arises from the electron pair correlation. 
Its existence is related to the breakdown of an 
approximate SU(2) particle-hole symmetry, from which 
an effective field theory of the new
excitations is constructed. 

\end{abstract}
\pacs{ 74.25.Jb, 71.27.+a, 71.45.-d}
\maketitle

It has become clear that the normal state of 
the high $T_c$ cuprate superconductors shows
many highly unusual properties. These properties 
are related to the fact that the high-$T_c$ cuprate
superconductors are doped antiferromagnetic (AF) 
Mott insulators. It is then not
surprising that the unusual behaviors are even more striking  in the
underdoped region, when the
concentration of doped holes is small.
A number of theoretical scenarios have emerged to understand the 
dynamics of the doped Mott insulators.
Anderson~\cite{Anderson:87} introduced the notion
of ``spin-charge separation'' in these materials and envisioned the
resonating valance bond  state --- a liquid of spin singlet pairs.  
This notion of spin-charge separation naturally accounts for all the
qualitative features of the spin
gap~\cite{PALee:98pre}. However, the spin-charge separation was 
proven only in one dimension whereas the cuprates are regarded as
quasi-two dimensional.  
On the other hand,  several scenarios of preformed pairs above
$T_c$~\cite{Emery-Kivelson:95,Randeria:97pre} have been proposed as
well to account for the underdoped (spin gap) regime of the high $T_c$
phase diagram.
In this paper, we  shall prove, based on the Hubbard model description
of the high $T_c$ cuprates, that 
there must appear a novel kind of  excitations 
that are spin singlet and carry charge $2e$ in the non-superconducting
state due to doping. Our results support the
ideas of spin-charge separation and preformed pairs. 
An effective field theory of these charge excitations 
will be derived.

Consider the extended Hubbard model in a 2D square lattice with 
spacing $a$, which we believe contains the essential physics of the
doped Mott insulators.
The Hamiltonian is
\begin{eqnarray}
H &= & -t \sum_{\avg{\Vx\Vx^\prime}\sigma} \left(c^\dag_{\sigma}(\Vx)
c_{\sigma}(\Vx^\prime) + h.c.\right)
+ J\sum_{\avg{\Vx\Vx^\prime}} \vec{S}(\Vx)\cdot \vec{S}(\Vx^\prime)
\nonumber \\
& &  +  U\sum_{\Vx} n_{\uparrow}(\Vx)n_{\downarrow}(\Vx) - \mu
\sum_{\Vx} n(\Vx) \,, 	\label{eq:HubbardH}
\end{eqnarray}
where $n_\sigma(\Vx)=c^\dag_{\sigma}(\Vx)c_{\sigma}(\Vx)$ with spin
$\sigma$ ($=\uparrow,\downarrow$), $n(\Vx)=n_\uparrow(\Vx)
+n_\downarrow(\Vx)$,
 and $\mu $ is the 
chemical potential that fixes the density of the system. 

The Hamiltonian (\ref{eq:HubbardH}) is invariant under the 
transformation, $c_\sigma \rightarrow {\sf U}_{\sigma\sigma^\prime}
c_{\sigma^\prime}$. This is
the usual SU(2) spin symmetry. Furthermore, 
near half-filling
the model has another (approximate)  SU(2) particle-hole symmetry,
sometimes called pseudospin
group~\cite{Affleck:89,Yang-Zhang:90,Schulz:90}.  
To see the symmetry, let us introduce two 
pseudospin doublets~\cite{Remark:SU(2)doublets}
$$\psi_{1}(\Vx)={c_{\uparrow}(\Vx) \choose (-)^\Vx 
c^\dag_{\downarrow}(\Vx)}\,, \qquad
\psi_{2}(\Vx)={c_{\downarrow}(\Vx) \choose -(-)^\Vx 
c^\dag_{\uparrow}(\Vx)}\, ,$$ 
where $(-)^\Vx\equiv e^{i\Vx\cdot \VQ}$ with $\VQ\equiv(\pi/a,\pi/a)$.
In terms of the doublets $\psi_{\alpha}$, the Hamiltonian (\ref{eq:HubbardH})
becomes
\begin{eqnarray}
H &=&
-t \sum_{\avg{\Vx\Vx^\prime}}
\psi^\dag_{\alpha}(\Vx)\psi_{\alpha}(\Vx^\prime) 
+J \sum_{\avg{\Vx\Vx^\prime}} \vec{S}(\Vx)\cdot \vec{S}(\Vx^\prime)
\nonumber \\
 && + {2U\over 3} \sum_{\Vx} \vec{\phi}\cdot\vec{\phi}
         + (U -2\mu) \sum_{\Vx}\phi_{3}
\,, \label{eq:H:SU(2)}
\end{eqnarray}
where 
$$\vec{S} = {1\over 4}\psi^\dag_{\alpha
l}\vec{\tau}_{\alpha\beta} \psi_{\beta l}$$ 
and 
$$
\vec{\phi} \equiv
{1\over 4}\psi^\dag_{\alpha
l}\vec{\tau}_{lm} \psi_{\alpha m}\,.
$$
The $\vec{\tau}$ is a
vector of the usual Pauli matrices.   (Note: Indices 
$\alpha,\beta,\cdots$ label two pseudospin doublets;
$l,m,\cdots$ label the rows of each.) 
Explicitly, $\phi_3={1\over 2}(n -1)$, and 
$(\phi_1,\phi_2) ={(-)^\Vx\over 2}(\Delta_- +\Delta_+,\, i\Delta_-
-i\Delta_+)$  where 
$\Delta_-(\Vx)\equiv c_\downarrow(\Vx)
c_\uparrow(\Vx)$ and $\Delta_+\equiv \Delta^\dag_-$ are the pairing
fields.   
The theory is thus obviously invariant at half filling 
($\mu=U/2$) under a global SU(2) pseudospin
transformation: 
$
\psi_{\alpha l} \rightarrow {\sf U}_{lm}\psi_{\alpha m}
$. Clearly, $\vec{\phi}$ transforms as a pseudospin vector.
The operators
\begin{equation}
J_{\pm} = \sum_\Vx (\phi_1 \pm i\phi_2)=\sum_\Vx
(-)^\Vx\Delta_{\pm}\,, \quad
J_3=\sum_\Vx \phi_3\,,
\end{equation}
generate the (approximate) SU(2)
pseudospin  symmetry, in the sense that
\begin{equation}
[H, J_\pm]= \pm(U-2\mu) J_\pm \,, \quad
[H, J_3]= 0 \, . \label{eq:[H,J]}
\end{equation}
The symmetry becomes exact at half filling with
$\mu=U/2$~\cite{SCZhang:90+91}. Note that $J$'s commute with the ordinary 
spin operators $\vec{S}$~\cite{Yang-Zhang:90}.

Consider the non-superconducting, underdoped regime of the phase
diagram.
Hole doping breaks the SU(2) pseudospin symmetry down to the ordinary
charge
U(1)$_{\rm c}$  subgroup in two ways.
At first, it breaks the 
symmetry {\em explicitly} at the Hamiltonian level. This is obvious from
Eq.~(\ref{eq:H:SU(2)}). 
Secondly, doping forces 
the ground state to `line' up in the $3$-direction (or
$\hat{z}$) in the pseudospin space while leaving the usual spin 
symmetry intact. This point is manifested by noticing
that doping induces a ground state such
that  $\avg{\phi_3}_0 
=-{\delta_h \over 2}\neq 0$ 
where  $\delta_h\equiv 1-\avg{n}_0$ is the concentration of
doped holes. 
The second way of breaking the symmetry is reminiscent 
of the notion of {\em spontaneous } symmetry breaking. 
One would immediately conclude from the Goldstone
theorem~\cite{note:GoldstoneTh} that  
the Goldstone modes must appear at low energy.
However, the present case   substantially differs 
from the case of spontaneously broken symmetries in that 
upon withdrawal of doping the ground state recovers the
full SU(2) pseudospin invariance~\cite{note:SSB}. 
The usual Goldstone theorem does not apply in the present case. We
need to carefully examine 
the breakdown of the pseudospin symmetry. 

Collective excitations, if any, can be identified by examing the
analytical properties of correlation functions. 
Consider the Fourier transform of
the following (retarded) electron-pair correlation function
\begin{equation}
D^{\mathrm{R}}_{-+}(\Vx-\Vx^\prime,t-t^\prime) = -i\theta(t-t^\prime)
 \avg{[\Delta_-(\Vx,t), \Delta_+(\Vx^\prime,t^\prime)]}_0 \,.
\label{eq:D(x,t)}
\end{equation}
At momentum $\Vk=\VQ$, 
\begin{equation}
D^{\mathrm{R}}_{-+}(\VQ,t-t^\prime) = -i\theta(t-t^\prime)
 \avg{[J_-(t), (-)^{\Vx^\prime}\Delta_+(\Vx^\prime,t^\prime)]}_0 \,,
\label{eq:D(Q,t)}
\end{equation}
where we have used the identity $J_-=\sum_\Vx
e^{-i\VQ\cdot\Vx}\Delta_-(\Vx)$. From (\ref{eq:[H,J]}), the time
dependence of $J_-$ can be determined via the equation of motion:
$$
J_-(t) = e^{-i(U-2\mu)(t-t^\prime)} J_-(t^\prime)\,.
$$ 
Also, we have
the equal-time commutator $[J_-, (-)^{\Vx^\prime}\Delta_+(\Vx^\prime)]
= -2\phi_3(\Vx^\prime)$. Therefore, we obtain exactly
\begin{equation}
D^{\mathrm{R}}_{-+}(\VQ,\omega) = -{2\avg{\phi_3}_0 \over
\omega-(U-2\mu) +i\eta}	\,. \label{eq:D(Q,omega)}
\end{equation}
The spectral function, defined in the Lehmann representation as 
\begin{equation}
A(\Vk,\omega) =\left({2\pi \over a}\right)^2 \sum_N
|\avg{0|\Delta_-(0)|N}|^2  
\delta^2(\Vk-\Vk_N)\delta(\omega-\omega_N) \,, \label{eq:A(k,omega)}
\end{equation}
can be determined exactly at momentum $\VQ$ from (\ref{eq:D(Q,omega)}):
\begin{eqnarray}
A(\Vk=\VQ,\omega) &=&-2 \Im
D^{\mathrm{R}}_{-+}(\VQ,\omega) \nonumber \\
&=& -4\avg{\phi_3}_0 \pi
\delta(\omega-(U-2\mu)) \,. \label{eq:A(Q,omega)}
\end{eqnarray}

Thus as long as the symmetry is broken by doping with 
$\avg{\phi_3}_0=-{\delta_h\over 2}\neq 0$, $A(\VQ,\omega)$
cannot vanish, but rather consists entirely of a term proportional to
$\delta(\omega-(U-2\mu))$. 
Such a term can obviously only arise in a theory that has particles of
spectrum  
$\omega=U-2\mu$ at $\Vk=\VQ$.
Furthermore, a delta function $\delta(\omega-(U-2\mu))$ can
only arise from single particle states; multi-particle states would 
contribute a continuum. The operator $\Delta_-$ is bosonic, 
so $\avg{0|\Delta_-|N}$ vanishes for
any fermion state $N$.  The state $\Delta^\dag_-|0\rangle$ is rotationally
invariant in spin space, so $\avg{0|\Delta_-|N}$  must vanish for any state
$N$ of non-zero spin. (We assumed a ground state of no long-range spin
order; an AF order will change our conclusion.)
Also  $\avg{0|\Delta_-|N}$ vanishes for any
state $N$ that has different (unbroken) internal quantum numbers 
from $\Delta_-$. A similar analysis beginning with the correlation function
$D^{\mathrm{R}}_{+-}$ in place of $D^{\mathrm{R}}_{-+}$ shall lead to
another collective mode of energy $-(U-2\mu)$ at $\Vk=\VQ$. 
We then conclude that there must exist two (spinless) 
bosonic collective
excitations of energy $\pm(U-2\mu)$ at momentum $\VQ$ and of the same
(unbroken) internal quantum numbers as $\Delta_-$ and $\Delta_+$, 
respectively.
  
Let us denote two  collective modes by
$W^-$ and $W^+$, respectively,  and  
consider how they should
transform under the unbroken U(1)$_{\rm c}$ symmetry. The quantum charge 
operator, if represented in terms of electron operators, is
$
Q= -e \sum_{\Vx} n = -2e(J_3 +{M\over 2})
$
where 
$M$ is the total number of sites of the lattice and the
electron charge is $-e$ in our convention.
Since $W^+$ ($W^-$)  must have the same
U(1)$_{\rm c}$ quantum number as $\Delta_+$ ($\Delta_-$), 
the commutation relation of $Q$ and $\zeta_\pm$ has to be identical with
that of $Q$ and $\Delta_\pm$.
Therefore,
\begin{equation}
[Q, W^\pm] = \mp 2e W^\pm \,.
\end{equation}
Physically, this means that ${W^-}$ and ${W^+}$ create
bosonic excitations of charges $2e$ and $-2e$,
respectively. Obviously  ${W^+}$ and ${W^-}$ are 
conjugate, so there exists in fact only one kind of such modes.
(Hereafter, we shall simply use $W$ to denote the collective
excitation.) 
We summarize these results into

\paragraph{Theorem.} If the ground state of the Hubbard model 
is such that $\avg{\phi_3}_0=-{\delta_h\over 2}\neq 0$ but 
$\avg{\Delta_\pm}_0= 0$,  
then there must exist a (spinless) {\it collective} excitation of 
charge $-2e$ and energy $|U-2\mu|$. 

We note that the above collective excitation may be viewed  
as a `pseudo-Goldstone-like' mode from  
the breakdown of the pseudospin symmetry. But we emphasize 
that the present case
conceptually differs from that of the spontaneously broken symmetry,
as discussed above (see also the note \cite{note:SSB}). 
Since $\mu=U/2$
when half-filled, we expect that the energy of the collective
excitation, 
$|U-2\mu|$,
is low for small $\delta_h$.
As we shall show
below from the view-point of effective field theory, 
the energy of this collective mode disperses quadratically away
from $\Vk=\VQ$. 

We would like to comment that the (spinless) 
collective excitation discussed in
the present paper is related to the {\it
triplet} of collective excitations of Zhang~\cite{SCZhang:90+91} and the
$\pi$ excitations (though of spin quantum number $1$) in the unified SO(5)
theory~\cite{SCZhang:97}. However, the triplet was proven to exist in
the {\it superconducting} state only, and Zhang~\cite{SCZhang:90+91}
argued that there is no direct way to couple to $J_+$ pair excitations
experimentally unless the ground state is superconducting.  By
contast, our collective excitation can physically show up
in the {\it
non-superconducting} state with $\delta_h$ acting as the role of order
parameter. For instance, 
Eq.~(\ref{eq:D(Q,omega)}) suggests that the collective excitation here should
be experimentally observable in the eletron pair response function 
at finite doping while the pairs need not to be condensed.  This is the main
result of the present paper.

It is quite tempting to find an expression of the state of the
collective excitation $W$  in
terms of electron states.
The spectral function (\ref{eq:A(k,omega)}) peaking 
at momentum $\VQ$ suggests that 
$\avg{\Delta_-(0)|W(\Vk)} \neq 0$. So a naive way of
looking for a state $|W(\Vk)\rangle$ is perhaps out of the states 
$\Delta^\dag_-(\Vx=0) |0\rangle$, projecting out
all  states of momenta different than $\Vk$.
We can then write the state $|W(\Vk)\rangle$ in
a general form
\begin{equation}
|W(\Vk)\rangle = 
\int_{\mathrm{BZ}} d^2 q \, 
f(\Vq) c^\dag_\downarrow(\Vk-\Vq) c^\dag_\uparrow(\Vq)
|0\rangle + \cdots \,. \label{eq:Wstate}
\end{equation}
$f(\Vq)$ is some function of $\Vq$ undetermined.
The terms indicated by `$\cdots$' represent all other possible
states of appropriate quantum numbers. 
So a simple $W$ boson may be viewed as a 
pair of electrons of opposite spins moving with a center-of-mass momentum
$\Vk$, and is not condensed in the normal state (i.e.,
$\avg{W}_0=0$). 
This is consistent with the picture of
preformed pairs in the non-superconducting 
state~\cite{Emery-Kivelson:95,Randeria:97pre}.
The formation of bound states of holes has been seen
numerically in a hole-doped antiferromagnet~\cite{Poilblanc++Dagotto:94}.

But one cannot simply identify the operator $W$ with $\Delta_-$. 
By a direct caculation,
\begin{eqnarray}
[\Delta_-(\Vx), \Delta^\dag_-(\Vx^\prime)]
& = & -2\phi_3(\Vx)\delta_{\Vx,\Vx^\prime} \nonumber \\
 &=& [\delta_h -\hat{n}^\prime(\Vx)] 
\delta_{\Vx,\Vx^\prime} \,,	\label{eq:[Delta, Delta]}
\end{eqnarray}
where $\hat{n}^\prime \equiv n -\avg{n}$. Therefore, $\Delta_-$ cannot
describe a true Boson since it fails yielding the standard
Bose-Einstein statistics, as required of  $W$ as a collective
excitation. Rather, Eq.~(\ref{eq:[Delta, Delta]}) implies that $W$
should describe
the collective motion of electron pairs with the quantum fluctuation
$\hat{n}^\prime$ integrated out. This point shall become a little more
manifest  in
the language of effective field theory.

An effective field theory of the $W$ collective excitation 
can be derived from
the breakdown of the pseudospin symmetry: 
SU(2)$\rightarrow$U(1)$_{\rm c}$ with $\avg{\phi_3}_0\neq 0$.
According
to the effective field theory approach of broken
symmetries~\cite{Weinberg:bk96:ch19+21},  
the (Goldstone-like) charge bosons may be identified (apart from
normalization) with  variables parameterizing the coset space
SU(2)/U(1)$_{\rm c}$.
We can write $\psi_\alpha$ as an SU(2)
transformation acting on $\tilde{\psi}_\alpha$: 
\begin{equation}
\psi_\alpha(\Vx,t) ={\sf U}(\Vx,t) \tilde{\psi}_\alpha(\Vx,t) 
\label{eq:psi->U*psi} \,, 
\end{equation}
where $\tilde{\psi}_\alpha$ is chosen such that 
$$\vec{\tilde{\phi}}={1\over 4} \tilde{\psi}^\dag_{\alpha} 
\vec{\tau} \tilde{\psi}_{\alpha} = (0,0,\sigma)\,.
$$
Accordingly, the $\vec{\phi}$ is expressed as
$\phi_{a}(\Vx,t) =
R_{a3}(\Vx,t)\sigma(\Vx,t) 
$
where $R_{ab}(\Vx,t)$ is a $3\times3$ orthogonal matrix determined by 
${\sf U}^{-1}\tau_a {\sf U} = R_{ab}\tau_b$.
Clearly,
$
\sigma(\Vx,t) =\sqrt{\sum_a \phi_a(\Vx,t)^2}$.
In place of the field variables $\psi_\alpha$, our variables now are
$\tilde{\psi}_\alpha$ and whatever other variables are needed to
parameterize the transformation ${\sf U}(\Vx,t)$.
We choose
\begin{equation}
{\sf U}(\Vx, t) = [1 -i \sum_{r=1}^2\tau_r
(-)^\Vx\tilde{\zeta}_r(\Vx,t)] 
/\sqrt{1+ \vec{\tilde{\zeta}}(\Vx, t)^2 }
\end{equation}
with 
\begin{eqnarray}
(-)^\Vx\tilde{\zeta}_1 &=& -\phi_2/( \phi_3+\sigma) \,, \\
(-)^\Vx\tilde{\zeta}_2 &= & \phi_1/(\phi_3+\sigma) \,.
\end{eqnarray}
The lattice
factor $(-)^\Vx$ is inherited from the definition of $\phi_r$ (see 
text following Eq.~(\ref{eq:H:SU(2)})) and is 
factorized out to ensure not to double the period of the lattice.
The field variables $\tilde{\zeta}_r$ can be combined into a complex
scalar field
$$
W= \tilde{\zeta}_1+ i\tilde{\zeta}_2\,,
$$
which may be identified as the collective excitation whose properties  
have been described in the theorem.

As an example, consider a simple case in which holes are uniformly
distributed:
$\avg{\sigma(\Vx)}_0=\avg{\phi_3(\Vx)}_0=-{\delta_h\over 2}$.
Also, we would generally 
expect a non-vanishing expectation value
of the valance bond field $\bar{\chi}_{\Vx\Vy}\equiv t\avg{
c^\dag_\sigma(\Vx)c_\sigma(\Vy)}_0$~\cite{Affleck-Marston:88}.
Starting with the Lagrangian corresponding to (\ref{eq:H:SU(2)}), 
we replace the $\psi_\alpha$
fields at each spacetime point with Eq.~(\ref{eq:psi->U*psi}) and
keep terms only relevant to the collective excitation field:
\begin{eqnarray}
L_W &=& -2\sigma \sum_\Vx {W^* \over \sqrt{1+|W|^2}} i\partial_t 
\left({W \over \sqrt{1+|W|^2}}\right)  \nonumber \\
&& 
+ \sum_{\avg{\Vx\Vy}} {\{{\chi}^*_{\Vx\Vy}[1-W^*(\Vx)W(\Vy)] +
c.c.\} \over \sqrt{[1+|W(\Vx)|^2][1+|W(\Vy)|^2]}}  \nonumber \\
& &  +\sigma(U-2\mu)\sum_\Vx {|W|^2 -1 \over |W|^2+1} \,.
\label{eq:L[W]:eff}
\end{eqnarray}
This
gives the effective theory of lower energy excitations about the
ground state we assumed.
Substituting all
$\sigma$ and $\chi$ fields with their 
expectation values and expanding the Lagrangian in powers of $W$ and
$W^*$, we have
\begin{eqnarray}
L_{\mathrm{eff}} &=& 
\delta_h \sum_\Vx W^* [i\partial_t -(U-2\mu)] W - \sum_{\avg{\Vx\Vy}}
2\Re\bar{\chi}_{\Vx\Vy} |W(\Vx)|^2  \nonumber \\
 && + \sum_{\avg{\Vx\Vy}}[ \bar{\chi}^*_{\Vx\Vy}
W^*(\Vx)W(\Vy) +c.c.] +\cdots \,. \label{eq:Lcb2}
\end{eqnarray}
The terms indicated by `$\cdots$' will contain high powers of the $W$
field. (A detailed justification of the above effective field theory will 
be given elsewhere.)
Different mean-field values of $\chi$ have been obtained for different
phases of the Heisenberg-Hubbard model~\cite{Affleck-Marston:88}.
For  a real uniform 
$\bar{\chi}_{\Vx\Vy}=\chi$, the Lagrangian (\ref{eq:Lcb2}) yields the
lowest order energy spectrum
$
E(\Vk) = {2\chi\over \delta_h} [2+\cos(k_xa) +\cos(k_ya)] +(U-2\mu)$.
Other phases of $\chi$ would yield different forms of energy spectrum but
all should have energy $U-2\mu$ at $({\pi/ a}, {\pi/ a})$.

In conclusion, we predicted a spinless 
collective excitation of charge $-2e$
in the
non-superconducting state of  the underdoped
high $T_c$ cuprates by analyzing the electron pair correlation.  
Identifying these new excitations thus offers an opportunity of
testing the validity of the Hubbard model for the high $T_c$ cuprates. 
Our result is consistent with 
the proposals of 
spin-charge separation~\cite{Anderson:87} and preformed 
pairs~\cite{Emery-Kivelson:95,Randeria:97pre}, which are widely used
to explain many important features of the normal state. 

I acknowledge my special intellectual debt to Professor Steven
Weinberg for very helpful discussions and correspondences. My special
thanks go to Eduardo Fradkin, Tony Leggett, Mike Stone, Kee-Su Park,
and J. M. Faundez Roman for stimulating discussions and criticisms.
The work was initialized at the University of Texas at Austin where it
was supported in part by NSF grant PHY-9511632 and the Robert A. Welch
Foundation, and is completed at the University of Illinois at
Urbana-Champaign where it is supported by NSF grant DMR-98-17941.


\begin{thebibliography}{10}
\bibitem{note:SSB}
\bibinfo{note}{A classic example of spontaneous symmetry breaking is the
  Heisenberg model of ferromagnetic spin interactions in which the spontaneous
  magnetization remains even after withdrawing the external magnetic field.}
\end{thebibliography}
\end{document}